\documentclass[doublecol]{epl2} 
\makeatletter
\newcommand\footnoteref[1]{\protected@xdef\@thefnmark{\ref{#1}}\@footnotemark}
\makeatother
\usepackage{graphicx}
\usepackage{dcolumn}
\usepackage{bm}
\usepackage{ulem}
\usepackage{xcolor}

\begin{document}

\title{Dynamical Casimir effects with atoms: from the emission of photon pairs to geometric phases}
\shorttitle{Dynamical Casimir effects with atoms} 

\author{authors}

\author{François Impens$^{1}$, Reinaldo de Melo e Souza$^{2}$, Guilherme C. Matos$^{1}$,  and Paulo A. Maia Neto$^{1,\footnote{pamn@if.ufrj.br}}$}


\institute{                    
  \inst{1} Instituto de F\'{i}sica, Universidade Federal do Rio de Janeiro,  Rio de Janeiro, RJ 21941-972, Brazil\\
\inst{2} Instituto de F\'{i}sica, Universidade Federal Fluminense, Niterói, RJ 24210-346,  Brazil}
\pacs{42.50.Ct}{Quantum description of interaction of light and matter; related experiments }
\pacs{03.75.Dg}{Atom and neutron interferometry}


\abstract{ 
The coupling between a moving ground-state atom and the quantum electromagnetic field is at the origin of several intriguing phenomena ranging from the dynamical Casimir emission of 
photons to Sagnac-like geometric phase shifts in atom interferometers. Recent progress in this emerging field
reveals unprecedented connections between
 non-trivial aspects of modern physics such as electrodynamic retardation, non-unitary evolution in open quantum systems, geometric phases, non-locality and inertia.
}

\maketitle

The interaction between matter and 
the quantum electromagnetic field is particularly intriguing when
motional effects play a major role. In the dynamical Casimir effect (DCE), 
photon pairs are emitted out of the vacuum state when neutral
 material surfaces are set into non-uniform motion (see \cite{Dalvit11,Dodonov20,Woods21,Munday21} for reviews). 
 
 DCE is usually considered within a macroscopic approach based on boundary conditions~\cite{Ford82,MaiaNeto96} 
 or scattering matrices~\cite{Jaekel92,Maghrebi13} for the quantum field. On a more fundamental microscopic level, DCE can be 
 described at the atomic scale~\cite{Souza18,Lo18,Belen2019,Kortkamp21,Fosco21} with the help of quantum optical Hamiltonian models for the atom-field coupling. 

 An analog of DCE with time-dependent boundary conditions led to an experimental demonstration in a superconducting waveguide~\cite{Wilson11}.
However, when considering optimal experimental conditions with moving atoms~\cite{Agusti21} or cavity mirrors~\cite{Dodonov92,Law94,Lambrecht96,Mundarain98,Plunien00,Crocce01,Dodonov11,Qin19}, 
the emission rates are still
too small to allow for a direct experimental verification even when resonance conditions are met. Typical orders of magnitude are also prohibitively small in the closely related quantum friction effect~\cite{Scheel09,Barton2010,Pieplow2013,Intravaia2016,Donaire2016,Reiche2020,Reiche2020B,Farias2020,Lombardo2021}. 

An alternative route is to track indirect motional signatures instead of the scarcely emitted photons or the tiny friction force.
In such context, moving atoms (rather than moving macroscopic particles)
near material surfaces
are ideally suited to probe DCE-like effects associated to the
atom-surface interaction, as the latter provides a rich playground in quantum electrodynamics~\cite{Laliotis21}. 
Time-dependent atom-surface forces were obtained when considering a non-stationary internal atomic state~\cite{Shresta2003,Vasile2008,Messina2010} or  transient conditions for the material surface~\cite{Behunin2011,Antezza2014}.
In this perspective paper, we 
review recent proposals and future directions in the emerging field of DCE with atoms, focusing on the time dependence resulting from the
atomic (center of mass) external motion. 
Naturally, probing 
DCE-like signatures in the non-relativistic regime remains a challenging task that requires high-precision devices. Atom interferometers~(AI) have
a successful track record of probing the atom-surface van der Waals (vdW) interaction potential~\cite{Perreault2005,Vigue09,Vigue11,Bender14}.
As shown in this perspective paper, AI are indeed natural candidates for measuring the tiny phase shifts arising from the motion of 
atoms in the vicinity of material surfaces, 
thus revealing intriguing DCE-like aspects
of the interaction with the quantum vacuum field. 

The motional correction of the phase associated to individual paths in the AI can be cast as a geometric phase~\cite{Pancharatnam56,Berry84} steered
by the atomic center of mass, which is treated as 
an external prescribed parameter in the Hamiltonian describing the atom-field interaction. In addition, 
the atomic motion leads to 
a nonlocal contribution associated to pairs of paths, which results from cross-talks between the interferometer paths mediated by the material surface. 
In some cases, such nonlocal contribution can be of the same order of magnitude of the single-path geometric phase correction. 
We discuss two examples where 
the motional phase corrections are of particular interest, and provide re-interpretations of previously published results that point to possible future directions in the field. 
We first review the microscopic dynamical Casimir
photon emission by a single 
ground-state atom. Such problem allows us to 
 introduce the concept of an interaction
 Hamiltonian parametrized by the position of the atomic center of mass, which we employ 
later to derive the motional atomic phases.

\textit{Microscopic dynamical Casimir effect.}
Within the electric dipole approximation, the atom-field interaction is described by the Hamiltonian (for alternative models and a more general discussion see \cite{thiru,andrews18,passante21})
\begin{equation}
\hat{V}(\mathbf{r}(t)) = -\mathbf{\hat{d}}^A\cdot\mathbf{\hat{E}}(\mathbf{r}(t)), \label{eq:dipolar potential}
\end{equation}
where
$\mathbf{\hat{d}}^A$ is the atomic dipole operator. 
As the  electric field operator $\mathbf{\hat{E}}$ is taken at the position $\mathbf{r}(t)$ of the atomic center of mass, the latter plays the role of a 
continuous and prescribed parameter of
the Hamiltonian $\hat{V}(\mathbf{r}(t)),$ which turns out to be 
 a key ingredient in all examples discussed in this paper. 
When considering moving atoms, it is in principle required
 to add the R\"ontgen correction~\cite{Baxter93,Wilkens94} to the dipolar Hamiltonian $\hat{V}(\mathbf{r}(t))$, so that the total interaction Hamiltonian reads
 $\hat{V}_R (\mathbf{r}(t))=\hat{V}(\mathbf{r}(t))- \mathbf{\hat{d}}^A\cdot\mathbf{v}(t)\times\mathbf{\hat{B}}(\mathbf{r}(t))$
 to order $v/c,$ where $\mathbf{v}(t)=\mathbf{\dot{r}}(t)$ and $\mathbf{\hat{B}}$ is the magnetic field. 
  
 We assume that the ground-state atom
   describes an harmonic motion with frequency $\omega_{\rm cm},$
a condition that can be met in one-dimensional optical dipole traps~\cite{Grimm00,Amico21}.
 The smallest internal transition frequency is typically orders of magnitude greater than $\omega_{\rm cm}$. Therefore, energy transfer from the motion goes into
the field channel only. In other words, motion induced excitation of internal states~\cite{Souza18, Belen2019} is negligible in the 
adiabatic limit,  and second-order parametric generation of dynamical Casimir photon pairs is dominant. The resulting microscopic DCE can be calculated to second-order of 
time-dependent perturbation theory from  $\hat{V}_R (\mathbf{r}(t))$~\cite{Kortkamp21}. Alternatively, one may calculate the DCE photon emission as a first-order effect 
of the effective Hamiltonian~\cite{Souza18}
\begin{eqnarray}
    \hat{H}_{\rm eff} (\mathbf{r}(t))=& -\frac{1}{2}\sum_{\mathbf{k}\lambda} \alpha (\omega_k)\left(\mathbf{\hat{E}}_{\mathbf{k\lambda}}+\mathbf{v}(t)\times\mathbf{\hat{B}}_{\mathbf{k\lambda}}\right)  \nonumber\\
    &\cdot \left(\mathbf{\hat{E}}+\mathbf{v}(t)\times\mathbf{\hat{B}}\right) , \label{heff}
\end{eqnarray}
where the sum is taken over the field modes 
  defined by wave vector ${\bf k}$ and polarization $\lambda$ and $ \alpha (\omega)$ is the atomic polarizability. The electric and magnetic field operators are calculated at 
  the atomic position $\mathbf{r}(t)$ as in (\ref{eq:dipolar potential}).
  
In order to demonstrate the equivalence between the descriptions by $ \hat{H}_{\rm eff}$ and by $\hat{V}_R (\mathbf{r}(t))$ as far as the field degrees of freedom are concerned, 
we start in the comoving frame where the atom is instantaneously at rest. 
In this case, the Hamiltonian assumes the form obtained in Ref.~\cite{Passante98}. 
When transforming to the 
laboratory frame, we neglect terms of the order $(v/c)^2$ or higher to obtain (\ref{heff}). 

Since $\hat{H}_{\rm eff} (\mathbf{r}(t))$ is valid only up to order $v/c$, 
the term involving the magnetic field squared in  (\ref{heff}) must be neglected for consistency. 
 $\hat{H}_{\rm eff} (\mathbf{r}(t))$
 operates only on field states as virtual internal transitions are already captured
  by the polarizability response function $\alpha(\omega) = \sum_e 2\omega_{eg}|\mathbf{d}^A_{eg}|^2/[3\hbar(\omega_{eg}^2-\omega^2)]$, where $g$ and $e$ denote the atomic ground and excited states, respectively, with $\omega_{eg}$ representing the corresponding transition frequencies
  and  $\mathbf{d}^A_{eg}=\langle e|\mathbf{\hat{d}}^A| g  \rangle $ the transition dipole matrix elements. 
  
   As $\hat{H}_{\rm eff} (\mathbf{r}(t))$ is quadratic in the field operators,
  it describes photon pair production at first order of perturbation theory. The results are identical to the second-order ones obtained from  $\hat{V}_R (\mathbf{r}(t)).$
On the other hand, higher-order contributions from the exact Hamiltonian $\hat{V}_R (\mathbf{r}(t))$ involve more than one atomic excited state, which is clearly not covered
  by the polarizability function  $\alpha(\omega).$ Thus, 
  the validity of  $\hat{H}_{\rm eff} (\mathbf{r}(t))$ is limited to first-order perturbation theory, which is precisely the one required 
  to calculate the DCE photon production rate. 
  
  Indeed, the probability for creation of the two-photon state $|1_{\mathbf{k_1}\lambda_1}1_{\mathbf{k_2}\lambda_2}\rangle$ is obtained from 
  $|\langle 0|\hat{H}_{\rm eff}|1_{\mathbf{k_1}\lambda_1}1_{\mathbf{k_2}\lambda_2}\rangle|^2.$ 
    In the rotating wave approximation and coarse-graining over a time scale much longer than $\omega_{\rm cm},$ 
    the photon pairs must satisfy energy conservation with  $c(k_1+k_2)=\omega_{\rm cm}.$
   We then evaluate the angular photon distribution 
  by adding over all all two-photon states compatible with energy conservation and containing a photon with the required energy and direction. 
   When considering a collection of atoms moving in phase so as to mimic the oscillation of a material planar surface, translation symmetry along any direction parallel to the surface also requires the conservation of
   momentum parallel to the surface, $\mathbf{k_1}_\parallel+\mathbf{k_2}_\parallel=0.$ In this case, the 
    qualitative features of the angular distribution obtained 
   from boundary conditions for a perfect-reflecting plane~\cite{MaiaNeto96} are re-obtained
   as a limiting case of 
a more general configuration representing spatiotemporal modulations of a surface~\cite{Kortkamp21}.

  We now focus on the total photon emission rate, which is obtained by integration of the angular and frequency spectra. 
  We write 
  the total photon emission rate in terms of the amplitude of oscillation $r_{\rm max}$ and of the peak velocity $v_{\rm max}=\omega_{\rm cm}r_{\rm max}.$
 The static polarizability is written in terms of a typical  atomic length scale $a$ as $\alpha(0)=4\pi\epsilon_0\,a^3,$
 where $\epsilon_0$ denotes the vacuum permittivity.
  We find
    a total emission rate given by~\cite{Souza18}
  $\Gamma = (23/5670\pi)(a/r_{\rm max})^6(v_{\max}/c)^8\omega_{\rm cm},$ which is clearly many orders of magnitude smaller than $\omega_{\rm cm}.$
 Given the minuteness of the photon emission effect, we look for alternative traces left by the atomic motion in the context of AI, as discussed in the next paragraph. 

\textit{Berry is late.}
 We consider below the AI configuration 
 used in Refs.~\cite{Perreault2005,Vigue09,Vigue11} to measure the quasi-static atom-surface vdW interaction:
 a two-path AI with one path propagating near a material surface, while the other path evolves far away and is thus immune to the surface interaction. 
Here, we focus on the  
DCE-like phase correction resulting from the relative motion between the propagating atoms and the surface. We provide below two complementary and equivalent interpretations of the motional phase correction, based either on the  concept of finite interaction time
or on a geometric argument \`a la Berry, respectively.

We first recall the expressions for the full vdW phase and for its quasi-static approximation.  The vdW phase acquired by an unpolarized atom evolving during the time $T$ can be written as a second-order Dyson term~\cite{Impens14} 
\begin{equation}
\phi = {\rm Re} \left[  \frac {{i}} {\hbar^2} \int_{-T/2}^{T/2} dt \int_{-T/2}^t dt'  \langle 0 | \hat{V}(\mathbf{r}(t),t) \hat{V}(\mathbf{r}(t'),t') | 0 \rangle \right]. \label{phiDyson}
\end{equation}
This expression is simplified by the fact that, in the considered AI configuration, only one path effectively interacts with the surface. When two or more paths propagate in the vicinity of the surface, one may also obtain non-local phases involving two paths at a time~\cite{Impens14,Impens13a,Impens13b}, as discussed later in this paper. 
 The position $\mathbf{r}(t)$ refers to the instantaneous average value of the position operator for the AI path near the surface, i.e. $\mathbf{r}(t) = \langle \hat{\mathbf{r}}(t) \rangle$. We have taken the dipolar Hamiltonian (\ref{eq:dipolar potential}) (in the interaction picture)
 as the contribution from the R\"ontgen correction to the motional phase shift is negligible in the examples discussed here.
 Apart from the atomic motion, we assume the system to be stationary and that the atom remains in its ground state, with $|0\rangle$ denoting
the combined atomic internal and (vacuum) field state. Generalizing to thermal states at arbitrary temperatures is straightforward. 
 Non-equilibrium configurations~\cite{Antezza2005,Obrecht2007,Buhmann2008,Behunin2010,Bartolo2016} can also be investigated along similar lines.

The quasi-static vdW phase is then obtained as a limiting case of 
eq.~(\ref{phiDyson}) 
by ``freezing'' the external atomic motion during the atom-surface interaction process mediated by a virtual photon. 
More precisely, the quasi-static result is derived by approximating $\mathbf{r}(t') \approx\mathbf{r}(t)$ in eq.~(\ref{phiDyson}), so that both dipolar Hamiltonians are taken at the same position. As a result, eq.~(\ref{phiDyson})  leads to the known quasi-static phase (see Ref.~\cite{Impens14} for a detailed derivation)
  \begin{equation}
  \label{eq:quasistatic}
  \phi^{{\rm qs}}= -\frac 1 \hbar \int_0^T dt\, U_{\rm vdW} (\mathbf{r}(t)), \label{eq:quasistatic}
  \end{equation}
  where $U_{\rm vdW} (\mathbf{r}(t)) $ is the van der Waals potential describing the atom-surface interaction at the instantaneous position $\mathbf{r}(t).$
  When approximating the total vdW phase (\ref{phiDyson}) by its quasi-static limit (\ref{eq:quasistatic}), we neglect the variation of the atomic position 
  during the time delay  $\tau=t-t'$ corresponding to the 
  elapsed time it takes for a virtual photon to propagate from the atom to the surface, scatter and propagate back. 
  For instance, in the case of a perfectly reflecting planar surface, 
  the delay is simply the round-trip light time $\tau=2z/c,$ with $z$ representing the atom-surface distance. More generally, real materials exhibit a finite electrodynamic 
  response time, making the total time delay $\tau$ longer than the round-trip light time. 
  
It is instructive to write the total vdW phase (\ref{phiDyson}) in terms of the time delay $\tau$ so as to emphasize its role.
In Ref.~\cite{Impens13b}, the total phase (\ref{phiDyson}) was recast in the form~(\ref{eq:quasistatic}) 
by substituting the instantaneous vdW potential $U_{\rm vdW} (\mathbf{r}(t))$ by its coarse-grained version defined  as follows:
$\overline{U}_{\rm vdW}(\mathbf{r}(t))= \frac {1} {\tau(t)} \int_{t}^{t+\tau(t)} dt' U_{\rm vdW}(\mathbf{r}(t'))$.
The coarse-graining
over the time delay $\tau$ captures 
 the intuitive notion that the whole set of atomic positions during the 
finite interaction time
 should be taken into account in order to estimate the atomic phase resulting from the atom-surface interaction, since the latter does not resolve an instantaneous atomic position.  The motional phase correction $\phi^{\rm mot}\equiv \phi- \phi^{{\rm qs}}$ is then written as 
\begin{equation}
\label{motionallocal}
\phi^{\rm mot}= -\frac 1 \hbar \int_0^T dt\, \left( \frac {} {} \overline{U}_{\rm vdW} (\mathbf{r}(t)) - U_{\rm vdW}(\mathbf{r}(t)) \right)  \label{eq:coarsegraining}
\end{equation}
 In the case of perfect reflectors considered in Refs.~\cite{Impens13a,Impens13b,Impens14}, the 
motional phase shift~(\ref{motionallocal}) is 
smaller than the quasi-static phase $\phi^{\rm qs}$ by multiplication by a factor
of the order of $v/c.$ 
Thus, $\phi^{\rm mot}$ is much smaller than the quasi-static phase for non-relativistic atoms.
On the other hand, one could expect that $\phi^{\rm mot}$ would be larger when considering real (dispersive) materials, since the time delay
$\tau$ also takes the medium finite response time into account. 

As an alternative method, one can re-derive the motional phase shift $\phi^{\rm mot}$ as the integral of a Berry connection along the atomic path,
 without any explicit reference to retardation
(for details see the supplementary material in Ref.~\cite{Matos21}). 
Indeed, as before we assume that
the atomic motion is too slow to excite internal states. As a result, the quantum state of the atom+field system follows adiabatically an instantaneous eigenstate
as the atom changes its position. 
Geometric phases in AIs have  been reported elsewhere~\cite{Miniatura92,Lepoutre12,Gillot13}, but not in the context of atom-surface interactions. 
Here, on the other hand, 
we focus on the shift $\phi^{\rm mot}$ arising from the motional correction of the atom-surface interaction. 
As before, we assume a small, localized atomic wave-packet following a prescribed average atomic trajectory given by $\mathbf{r}(t)$, which plays the role of an external parameter steering the dipolar Hamiltonian $\hat{V}(\mathbf{r}(t))$. 
 We introduce the free Hamiltonian eigenstates basis  $|n \rangle$ satisfying $\hat{H}_0 | n \rangle = \hbar \omega_n |n \rangle.$  Note that $n$ actually corresponds to a continuous parameter associated to a wave-vector, an electric field polarization and an internal atomic state. As in the discussion of the local phases, we assume the initial
 state $|0\rangle$ to be the ground state of $H_0,$ which correspond  to the internal atomic ground state in the quantum vacuum field. 
 More general initial states can be considered along similar lines. 

We start from eq.~(\ref{phiDyson}) and take the Taylor expansion $\mathbf{r}(t') \simeq \mathbf{r}(t) - (t-t')\, {\mathbf{v}}(t)$ in order to follow
 the atomic position during the time delay $\tau=t-t'$. The motional phase shift corresponds to the term proportional to the velocity. 
After changing the integration variables to $\mathbf{r}(t)$ and $\tau$ and expanding the dipolar Hamiltonian $\hat{V}(\mathbf{r}(t'),t')$, we find
\begin{equation}
\phi^{\rm mot}= \frac {i} {{2} \hbar^2} \int_{{\cal P}} d\mathbf{r}\cdot \int_{0}^{T} d \tau  \: \tau \:  \langle 0 | \hat{V}(\mathbf{r},t) \nabla_{\mathbf{r}} \hat{V}(\mathbf{r},t-\tau) | 0 \rangle   \label{phinabla}
\end{equation}
where the time $t(\mathbf{r})$ is obtained from the inverse of the function $\mathbf{r}(t)$ as the atom follows path ${{\cal P}}.$

We now assume that the interaction time is sufficiently long to resolve the frequency scales associated to the spectrum of $H_0:$
 $T \gg \omega_n^{-1}.$ Then, using the completeness relation for the states  $ | n \rangle,$ we re-write (\ref{phinabla}) as 
\begin{eqnarray}
\label{eq:motional_correction_final}
 \phi^{\rm mot}   =    \frac {i} {2 \hbar^2} \int_{{\cal P}} d \mathbf{r}\cdot   \left( \sum_{n \neq 0} \frac { \langle 0 | \hat{V}(\mathbf{r}) | n \rangle \langle n |  \nabla_{\mathbf{r}} \hat{V}(\mathbf{r}) | 0 \rangle} {(\omega_n-\omega_0)^2} \right. \nonumber \\   
 -  \left.  \frac {\langle 0 |  \nabla_{\mathbf{r}} \hat{V}(\mathbf{r}) | n \rangle \langle n |  \hat{V}(\mathbf{r}) | 0 \rangle}
 {(\omega_n-\omega_0)^2}   \right)  \label{eq:motional2}
\end{eqnarray}
 
In order to show that the motional phase~(\ref{eq:motional2}) is a geometric
phase, we interpret the atomic center of mass $\mathbf{r}$ as a continuous parameter on which depends the instantaneous dressed state $| \psi_0( \mathbf{r} ) \rangle$ of the 
atom+field system when taking the interaction into account. 
 Up to second order of stationary perturbation theory, $| \psi_0( \mathbf{r} ) \rangle$ is such that (\ref{eq:motional2}) can be cast as the integral of a Berry connection~\cite{Shapere1989}
of the form 
\begin{equation}\label{geometricphase}
\phi^{\rm mot}   =  i \int_{{\cal P}} d \mathbf{r}\cdot \langle \psi_0(\mathbf{r}) | \nabla_{\mathbf{r}} \psi_0(\mathbf{r}) \rangle.
\end{equation}
As in the discussion of the photon emission effect, we have assumed that the atomic motion is too slow to excite internal atomic transitions, 
and then the  atom+field quantum state follows adiabatically the dressed state $ | \psi_0( \mathbf{r} ) \rangle$ as $\mathbf{r}$ changes in time. Thus, 
we were able to recast (\ref{eq:motional2}), which was derived from the  time-dependent perturbation result (\ref{phiDyson}), in terms of a geometric phase 
obtained within stationary perturbation theory. 

In short, we have obtained two equivalent forms for
the DCE-like  motional phase that corrects the 
 quasi-static result (\ref{eq:quasistatic}). 
 The motional phase shift $\phi^{\rm mot}$ can be seen as resulting from the finiteness of the  interaction time, 
 which is captured by the coarse-graining indicated by (\ref{eq:coarsegraining}).
 Alternatively, $\phi^{\rm mot}$ can also be cast as a geometric phase (\ref{geometricphase})
 as the atomic position drives the full atom+field system along an adiabatic quantum trajectory.

\textit{Non-local geometric vdW phases.} Here, we assume that the atomic beam-splitter employed in the AI arrangement is such as to preserve 
the internal ground state, as for instance in the already mentioned vdW experiments with gratings~\cite{Perreault2005,Vigue09,Vigue11}.
In this case, when two or more AI paths are close to the material surface, 
the motional correction leads to a nonlocal phase shift~~\cite{Impens13a,Impens13b,Impens14}
 in addition to (and in some cases with the same order of magnitude of) the local one 
 $\phi^{\rm mot}$ discussed above. 

Precisely, the total relative phase measured in the atom interferometer with two paths is written as
\begin{equation}
  \Delta  \phi = \phi_1-\phi_2 + \phi_{12} \, ,
  \label{phasediff}
\end{equation}
where $\phi_j$ ($j=1,2$) is the local vdW phase for path $j$ discussed before.
 The nonlocal quantum phase $\phi_{12}$ depends simultaneously on the pair $(1,2)$ of distinct AI paths in a non-separable manner. 
 This contrasts with the usual paradigm of atom interferometry for which each phase is attached to a single path. 
Indeed,
most atomic phases encountered in AIs 
are not only local but also dynamical, i.e., they are 
obtained from the integration of a potential along a single path as in the case of the vdW quasi-static phase (\ref{eq:quasistatic}). Typically, 
such local dynamical phases correspond
to the average trajectory in the case of sufficiently narrow wave-packets or for Gaussian wave-packets evolving in quadratic potentials~\cite{Borde01,Impens09}. 

 It is illustrative to compare the non-local phase with the quantum Cheshire cat effect~\cite{Aharonov13,Das20,Liu20,CeshireCat21}. In this paradigm, a physical property (the ``grin'',  here the internal atomic dofs) associated to a quantum particle
is spatially disconnected from the particle location and can be isolated from the carrier (the ``cat'', here the external atomic dof). In our configuration, the ``Cheshire cat'' has a common head (the internal atomic dof) and two distinct legs (the external dofs). The non-locality of the internal atomic dofs allows for cross-talks between the two AI paths
which lie at the origin of the nonlocal phase. By the same token, nonlocal terms contribute
to decoherence by spontaneous emission of an excited atom in a quantum superposition of external states~\cite{Souza2016}. 

\begin{figure}[htbp]
\begin{center}
\includegraphics[width=4.cm]{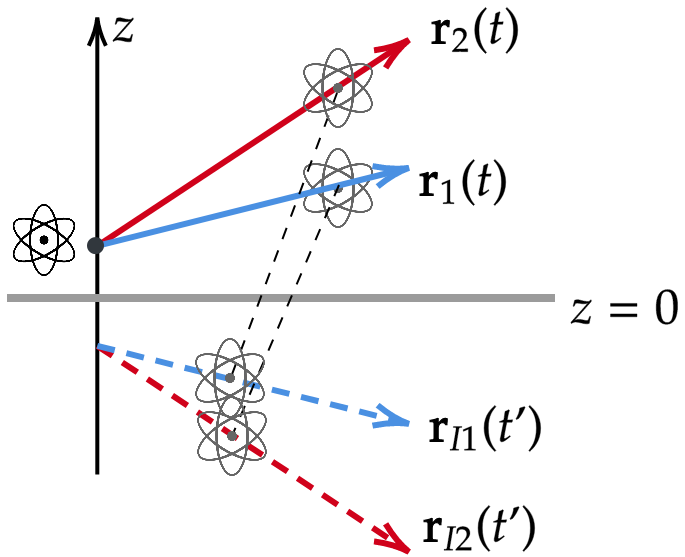}
\end{center}  \caption{(Color
  online) Diagrammatic representation of the nonlocal phase as a result of cross-talks between the AI paths $\mathbf{r}_1(t)$ and $\mathbf{r}_2(t)$ near a perfectly-reflecting planar surface at $z=0.$ The dashed lines with arrows represent the image paths $\mathbf{r}_{I1}(t)$ and $\mathbf{r}_{I2}(t).$ Each diagram involves a given  time $t$ on one path and a retarded time $t'$ on the image of the other path. The non-local phase $\phi_{12}$ arises 
  as the difference between the contributions associated to
these two diagrams.}
 \label{Fig1}
\label{fig:diagrams}
\end{figure}

The cross-talks can be understood in terms of the standard picture of the vdW atom-surface interaction as resulting from the 
interaction between the fluctuating atomic dipole and its image representing the induced charge on the material surface~\cite{Laliotis21}. 
When the atomic external state is split into two wavepackets 
associated to the same internal state, each packet interacts not only with its own image but also with the image of the other packet, as illustrated by
fig.~\ref{fig:diagrams}. Such cross-interactions do not contribute in the quasi-static limit because $\phi_{12}$ arises as the difference between the two resulting cross-talks
shown in the figure. It is in fact the motion with respect to the surface that breaks the symmetry 
between the two cross-interaction terms, thus leading to a nonzero motional nonlocal phase, as discussed below. 

More specifically,  the nonlocal phase is obtained
by treating the external atomic dof as an open quantum system with the internal atomic dipole and electric field dofs constituting the quantum environment.
The atomic external, internal and field dofs are coupled through the
dipolar Hamiltonian $\hat{V}(\hat{\mathbf{r}}_A)= -\hat{\mathbf{d}}^A \cdot \hat{\mathbf{E}}(\hat{\mathbf{r}}_A),$
which differs from (\ref{eq:dipolar potential}) by considering the atomic center of mass position $\hat{\mathbf{r}}_A$
 as a quantum operator rather than a prescribed function of time. We trace out the environmnent dofs to obtain the reduced density operator for the center of mass to second order
 in the dipolar interaction. The environment footprint is obtained either as an influence functional~\cite{Impens13a} by employing the Feynmann-Vernon path integral
 formalism~\cite{Feynmann63} or as a complex phase within standard second-order time-dependent perturbation theory~\cite{Impens14}.

For concreteness, we consider a planar perfectly-reflecting surface placed at $z=0,$
and an AI with two paths flying near its close vicinity
with the same velocity component parallel to the surface (see fig.~\ref{Fig1}). 
For a two-level atom, 
the resulting non-local phase shift $ \phi_{12}$ is then given by the 
geometric integral~\cite{Impens13b}
 \begin{equation}\label{eq:double path phase}
  \phi_{12} = \frac{3\, \omega_0\alpha(0)}{4\pi\epsilon_0c} \, \int_{-T/2}^{T/2} dt\, \frac{\dot{z}_1(t)-\dot{z}_2(t)}{(z_1(t)+z_2(t))^3},
  \label{eq:doublepathgeometric}
  \end{equation}
where $\alpha(0)$ and $\omega_0$ denote the static atomic polarizability
 and the  transition frequency, respectively. 
When deriving (\ref{eq:doublepathgeometric}), we assumed very narrow wave-packets around their average positions $\mathbf{r}_j(t),\, j=1,2,$ with $z_j(t)$ representing the  axial components. 
 A more general result valid for finite-width wave-packets can be found in Ref.~\cite{Impens14}. 
 
 As discussed in connection with fig.~\ref{Fig1}, the nonlocal phase $\phi_{12}$
arises as the difference between the cross interactions between each path and the image of the other one. 
Because of the electrodynamic time delay $\tau,$ each 
 cross interaction involves a given time $t$ and its corresponding retarded one $t'=t-\tau.$ As the image paths lag behind the real ones by different 
 amounts depending on the value of $\dot{z}_j,$ the resulting subtraction leads to the nonlocal phase (\ref{eq:double path phase})  depending on the 
 difference between the velocity components along the $z$-axis.
Thus, it is the relative motion with respect to the surface that sets the sign and magnitude of the nonlocal phase shift, which 
can be isolated from local contributions by seeking a violation of phase additivity in an AI with three or more paths~\cite{Impens13b}. 
 
As a final remark, we note that the phase~(\ref{eq:double path phase}) fulfills two key features of geometric phases~\cite{Lepoutre12}: 
 first, it does not depend on the magnitude of the wave-packets' velocities (after time integration); 
 and, second,  it changes sign when the direction of propagation is reversed. 
 This is in contrast with dynamical phases, which are inversely proportional to the  
 magnitude of the velocity 
 and insensitive to the direction of propagation. 
 The motional phase~(\ref{eq:double path phase}) is thus a non-local geometric phase, and the only example of this kind reported so far to our knowledge.
   
\begin{figure}[htbp]
\begin{center}
\includegraphics[width=4.2cm]{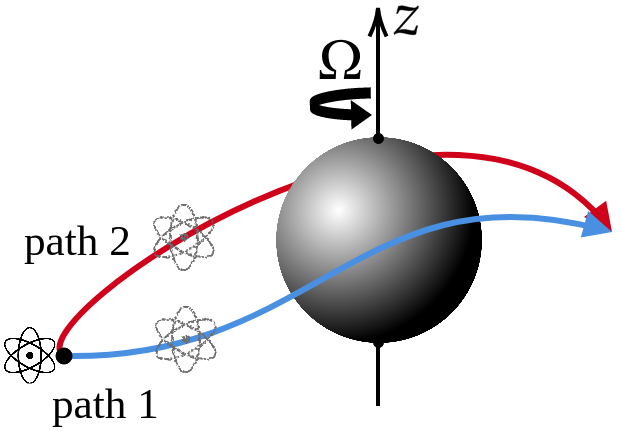}
\end{center}  \caption{(Color
  online) Quantum Sagnac effect: atom interferometer 
  with a spinning particle (angular velocity $\Omega$) placed in between its arms.}
  \label{fig2}
 \end{figure}

\textit{Quantum Sagnac effect.} We now apply the concepts of local and nonlocal motional phases to an example of particular interest. 
A spherical particle of radius $a$ spinning at a constant angular velocity $\boldsymbol{\Omega}$ is placed in between the arms of the AI, as illustrated by fig.~\ref{fig2}.
The resulting motional vdW phase contains a $\Omega-$dependent term which is reminiscent of the Sagnac effect, with the remarkable difference that here the rotation is confined to a limited
region of space. 
As before, we assume the atom to be in its ground state $|g\rangle.$
A full quantum approach was recently employed to derive explicit results for a two-level atomic model~\cite{Matos21}. Here, we present an alternative derivation for a more general
multi-level atom but 
restricted to short distances satisfying the condition that the light  travel time between particle and atom  is much smaller than both the atomic and  particle response times.
 In this case, electrodynamic retardation is negligible as far as the internal dofs are concerned.
 We then
neglect field fluctuations and follow a semi-classical approach without considering the quantum electromagnetic field. We also assume that the spinning particle is sufficiently small to be approximated by its electric dipole moment $\hat{\mathbf{d}}^S(t).$

The atom-particle interaction is then described by the instantaneous dipole-dipole interaction Hamiltonian
\begin{equation}
     \hat{V}_{\rm dd} (\mathbf{r}(t),t) = \frac{\hat{\mathbf{d}}^A(t)\cdot\hat{\mathbf{d}}^S(t)-3(\hat{\mathbf{d}}^A(t)\cdot{\mathbf{u}_{r}})(\hat{\mathbf{d}}^S(t)\cdot\mathbf{u}_{r})}{4\pi\varepsilon_0\,r^3} \, , \label{vdipdip}
\end{equation}
where $\mathbf{u}_{r}$ is the unit radial vector with 
 $\mathbf{r}(t)=r(t)\mathbf{u}_{r}(t)$ denoting the position of the atom with respect to the sphere's center at time $t.$  
 
 The local motional phase acquired by the atom during the interaction time $T$ can be obtained by
 replacing the atom-field Hamiltonian $  \hat{V}$ by the dipole-dipole one (\ref{vdipdip}) in 
 eq.~(\ref{phinabla}), where now the complete ground state reads $|0\rangle = |g\rangle |0\rangle_S,$ with 
 $|0\rangle_S$ representing the ground state of the particle's internal dofs.  
 By developping eq.~(\ref{phinabla}), we find contributions of 
 the form  
 ($i,j,m,n$ denoting 
 Cartesian components)
 $ {}_S\langle 0| d^S_i(t) d^S_j(t-\tau)|0\rangle_S \langle g| d^A_m(t) d^A_n(t-\tau)|g\rangle$. Since the atom is isotropic,  its dipole correlation function is symmetric under the interchange $i\leftrightarrow j$. 
As a consequence, the $\Omega-$dependent part of the motional phase for the AI path $k$ simplifies to 
\begin{eqnarray}
    \phi_k^\Omega &=& \frac{9}{(4\pi\varepsilon_0)^2\hbar^2}\int_{{\cal P}_k} dr_i\frac{r_j}{r^8} \int_0^T d\tau\, \tau\, {}_S\langle 0| \hat{d}^S_j(t) \hat{d}^S_i(t-\tau)|0\rangle_S
     \nonumber
      \\
    &&\times\,\langle g| (\hat{\boldsymbol{d}}^A(t)\cdot\mathbf{u}_{r})(\hat{\boldsymbol{d}}^A(t-\tau)\cdot\mathbf{u}_{r})|g\rangle,\label{phiOmega}
\end{eqnarray}
where $t(\mathbf{r}_k)$ is determined as the inverse of the function $\mathbf{r}_k(t)$ corresponding to path ${{\cal P}_k}.$
We have employed Einstein notation for the sum over Cartesian components. We find
\begin{eqnarray}
    \phi_k^\Omega &=& \frac{3}{(4\pi\varepsilon_0)^2\hbar^2}\sum_e |\mathbf{d}_{eg}^{A}|^2\int_{{\cal P}_k} dr_j\frac{r_i}{r^8}  \cr\cr 
    &&\times\int_0^T d\tau \tau  e^{-i\omega_{eg}\tau} {}_S\langle 0| \hat{d}^S_i(t) \hat{d}^S_j(t-\tau)|0\rangle_S  \, . \label{phiOmega2} \,
\end{eqnarray}
When deriving  (\ref{phiOmega2}) from (\ref{phiOmega}), we have used that 
only diagonal terms of the atomic dipole correlation function suvive after summing over all excited states
as required by isotropy.

We now write the particle's dipole correlation function as 
${}_S\langle 0| \hat{d}^S_i(t) \hat{d}^S_j(t-\tau)|0\rangle_S =\frac12[ \hbar({\tilde \alpha}_{ij}(\tau)-{\tilde \alpha}_{ji}(-\tau)) +
{}_S\langle 0| \{\hat{d}^S_i(t), \hat{d}^S_j(t-\tau)\}|0\rangle_S],$
where ${\tilde \alpha}_{ij}(\tau)$ are the components of the spinning particle's polarizability tensor in the time domain. 
The symmetric part of
the correlation function is not modified by the spinning and hence does not contribute to the $\Omega-$dependent phase shift. 
In the frequency domain, the modification of the polarizability tensor due to non-relativistic rotation
 is $\alpha_{ij}^{\boldsymbol{\Omega}}(\omega) = i\alpha_S'(\omega)\epsilon_{ijk}\Omega_k$~\cite{Manjavacas10},
 where $\alpha_S(\omega)$ is the polarizability of the spherical particle at rest. 
 Combining these results and integrating (\ref{phiOmega})  by parts, 
we obtain our final result
\begin{equation}
    \phi_k^{{\Omega}} = \sum_e \frac{3|\mathbf{d}_{eg}^{A}|^2{\rm Re}\, \alpha_S''(\omega_{eg})}{(4\pi\varepsilon_0)^2\hbar}\int_{{\cal P}_k} \frac{d\boldsymbol{r}\cdot \boldsymbol{\Omega}\times\boldsymbol{r}}{r^8}  . \label{sagnacsp}
\end{equation}
As in the standard Sagnac effect, the quantum Sagnac local phase (\ref{sagnacsp})
corresponds to a geometric phase given by a line integral of a vector potential proportional to the sphere's angular velocity $\Omega$,
which brings an analogy with the Aharonov-Bohm effect~\cite{Aharonov1959} into play~\cite{Dalibard11, Paiva21}. 

A realistic example is to take atomic trajectories along straight lines~\cite{Perreault2005, Vigue09, Vigue11} on a plane perpendicular to ${\bf \Omega}=\Omega \mathbf{u}_z$, with
$\mathbf{r}_k(t) = vt\,\mathbf{u}_x+y_k\,\mathbf{u}_y.$ 
Eq.~(\ref{sagnacsp}) then leads to 
    $\phi_k^{{\Omega}} = \frac{15\pi}{16}\left(\frac{\ell_\Omega}{y_k}\right)^6{\rm sgn}(y_k)$, 
where ${\rm sgn}$ is the sign function and $\ell_\Omega$ stands for the characteristic length scale $\ell_\Omega \equiv (\sum_e |\mathbf{d}_{ge}^{A}|^2{\rm Re} \,\, \alpha_s''(\omega_{ge})\Omega/(4\pi\varepsilon_0)^2\hbar)^{1/6}$. 
 If one of the paths is much closer to the sphere than the other, 
the total phase difference (\ref{phasediff}) is simply the difference between the local Sagnac phase shifts.
On the other hand, 
when $y_2\sim y_1$ the non-local phase shift becomes relevant. For instance, when $y_2=-y_1$ 
the non-local shift reduces the total phase by about one third in the case of a two-level atomic model:
$
    \Delta\phi = \frac{21\pi}{16}\left(\frac{\ell_\Omega}{y_1}\right)^6
$
(see~\cite{Matos21} for details). 

{\it Conclusion.} 
The emission of photon pairs out of the vacuum field by accelerated ground-state atoms is an intriguing prediction of quantum theory. 
DCE-like extensions in the field of atom-surface interactions give rise to new phenomena which can in principle be probed by atom interferometers. 
Indeed, motional corrections of the quasi-static phase shift include not only a coarse-graining of the atom-surface distance due to the finite interaction time, but also to a non-local phase shift. The latter cannot be decomposed as a difference between quantities associated to individual interferometric paths. Instead, it belongs to pairs of paths, as 
it arises from the interaction between one wave-packet and the surface image of the other one. 
When considering two paths with comparable distances to the surface, 
the nonlocal and local motional phase shifts have the same order of magnitude. 

In contrast to the quasi-static phase, the motional phase shifts are of a geometric nature.
An analogy with the Aharanov-Bohm  effect is particularly clear when considering the interaction with a spinning particle placed in between the 
interferometer paths. The resulting quantum Sagnac phase appears as the line integral of an effective vector potential proportional to the particle's angular velocity.
While the standard Sagnac phase results from rotating the entire frame of reference, its quantum version arises instead from the rotation of a single particle.
 In a sense, the quantum Sagnac effect relates to Mach's ideas on how the concept of  ``inertiality" connects to motion in space~\cite{Sciama1953}. 
 
Given their stronger coupling with the field, quantum emitters such as quantum dots or NV centers in diamond~\cite{Lombardo2021}
are candidates to replace the role of atoms, particularly
as the cooling of nanoparticles' external dofs reach the quantum level~\cite{Delic2020,Tebbenjohanns2021,Stickler2021}. 
The emerging field of DCE with atoms or quantum emitters
 has the potential to reveal a plethora of new phenomena
at the intersection between quantum nonlocality, geometric phases, and inertial effects.

\acknowledgments

We thank 
Ryan Behunin, Claudio Ccappa Ttira,
 Jean Dalibard, Ricardo Decca, Manuel Donaire, Antonio Khoury, and Marcelo da Silva Neto for discussions.
We acknowledge funding from the Brazilian agencies 
 Conselho Nacional de Desenvolvimento Cient\'{\i}fico e Tecnol\'ogico (CNPq),
Coordena\c c\~ao de Aperfei\c coamento de Pessoal de N\'{\i}vel Superior
 (CAPES), 
 Instituto Nacional de Ci\^encia e Tecnologia de Informa\c c\~ao Qu\^antica (INCT-IQ),
 Instituto Nacional de Ci\^encia e Tecnologia de Fluidos Complexos (INCT-FCx),
 and the
Research Foundations of the States of Rio de Janeiro (FAPERJ) (202.874/2017, 210.242/2018 and 210.296/2019) and S\~ao Paulo (FAPESP)(2014/50983-3).


\begin{thebibliography}{0}
\expandafter\ifx\csname
natexlab\endcsname\relax\def\natexlab#1{#1}\fi
\expandafter\ifx\csname bibnamefont\endcsname\relax
  \def\bibnamefont#1{#1}\fi
\expandafter\ifx\csname bibfnamefont\endcsname\relax
  \def\bibfnamefont#1{#1}\fi
\expandafter\ifx\csname citenamefont\endcsname\relax
  \def\citenamefont#1{#1}\fi
\expandafter\ifx\csname url\endcsname\relax
  \def\url#1{\texttt{#1}}\fi
\expandafter\ifx\csname
urlprefix\endcsname\relax\def\urlprefix{URL }\fi
\providecommand{\bibinfo}[2]{#2}
\providecommand{\eprint}[2][]{\url{#2}}


\bibitem{Dalvit11}
DALVIT  D. A. R., MAIA NETO P. A.  and MAZZITELLI F. D., {\it Lect. Notes Phys.}, \textbf{834} (2011) 287.

\bibitem{Dodonov20}  DODONOV V. V.,
\textit{Physics},  {\bf 2} (2020) 67.

\bibitem{Munday21}  GONG T. \textit{et al.},
 \textit{Nanophotonics}, {\bf 10} (2021) 523.

\bibitem{Woods21}
 WOODS L. M., KR\"UGER M. and  DODONOV V. V., 
\textit{Appl. Sci.}, {\bf 11} (2021) 293.	

\bibitem{Ford82} FORD L. H. and VILENKIN  A., {\it Phys. Rev. D}, {\bf 25} (1982) 2569. 

\bibitem{MaiaNeto96} MAIA NETO P. A. and MACHADO L. A. S., {\it Phys. Rev. A}, {\bf 54} (1996) 3420.

\bibitem{Jaekel92} JAEKEL M. T. and REYNAUD S., \textit{Quantum Optics}, {\bf 4} (1992) 39. 

\bibitem{Maghrebi13} MAGHREBI M. F., GOLESTANIAN R. and KARDAR M.,  \textit{Phys. Rev. D}, {\bf 87} (2013) 025016.

\bibitem{Souza18} DE MELO E SOUZA R., IMPENS F. and MAIA NETO P. A., \textit{Phys. Rev. A}, {\bf 97} (2018) 032514.

\bibitem{Belen2019} BEL\'EN FARIAS M. \textit{et al.}, \textit{Phys. Rev. D}, {\bf 100} (2019) 036013.

\bibitem{Lo18} LO L. and LAW C. K., \textit{Phys. Rev. A}, {\bf 98} (2018)  063807. 

\bibitem{Fosco21} FOSCO C. D., LOMBARDO F. C, and MAZZITELLI F. D.,  \textit{Universe}, {\bf 7} (2021) 158.

\bibitem{Kortkamp21}
DALVIT D. A. R. and KORT-KAMP W. J. M.,
\textit{Universe}, {\bf 7} (2021) 189.

\bibitem{Wilson11} WILSON C. M. \textit{et al.}, \textit{Nature}, {\bf 479} (2011) 376.

\bibitem{Agusti21} AGUST\'I A. \textit{et al.}, \textit{Phys. Rev. A}, {\bf 103} (2021) 062201.  

\bibitem{Dodonov92} DODONOV V. V. and KLIMOV A. B., \textit{Phys. Lett. A}, {\bf 167} (1992) 309.

\bibitem{Law94} LAW C. K., \textit{Phys. Rev. A}, {\bf 49} (1994) 433.

\bibitem{Lambrecht96} LAMBRECHT A.,  JAEKEL M. -T. and REYNAUD S., \textit{Phys. Rev. Lett.}, {\bf 77} (1996) 615.

\bibitem{Mundarain98} MUNDARAIN D. F. and MAIA NETO P. A., \textit{Phys. Rev. A}, {\bf 57} (1998) 1379. 

\bibitem{Plunien00} PLUNIEN G., SCH\"UTZHOLD R., and SOFF G., \textit{Phys. Rev. Lett.}, {\bf 84} (2000) 1882.

\bibitem{Crocce01} CROCCE M., DALVIT D. A. R. and MAZZITELLI F. D., \textit{Phys. Rev. A}, {\bf 64} (2001) 013808.

\bibitem{Dodonov11} DODONOV A. V. and DODONOV V. V., \textit{Phys. Rev. A}, {\bf 85} (2012) 015805.

\bibitem{Qin19} QIN W., \textit{et al.}, \textit{Phys. Rev. A}, {\bf 100} (2019) 062501.

\bibitem{Scheel09} SCHEEL S. and BUHMANN S. Y., \textit{Phys. Rev. A}, \textbf{80} (2009) 042902.

\bibitem{Barton2010} BARTON G.,  \textit{New J. Phys.}, {\bf 12} (2010)  113045.

\bibitem{Pieplow2013} PIEPLOW G. and HENKEL C., \textit{New J. Phys.}, {\bf 15} (2013) 023027.

\bibitem{Intravaia2016} INTRAVAIA F. \textit{et al.}, \textit{Phys. Rev. Lett.}, {\bf  117} (2016) 100402.

\bibitem{Donaire2016}  DONAIRE M. and LAMBRECHT A., \textit{Phys. Rev. A}, \textbf{93} (2016) 022701.

\bibitem{Reiche2020} REICHE D. \textit{et al.}, \textit{Phys. Rev. A}, {\bf 102} (2020) 050203(R). 

\bibitem{Reiche2020B} REICHE D., BUSCH K. and INTRAVAIA F., \textit{Phys. Rev. Lett.}, {\bf 124} (2020) 193603. 

\bibitem{Farias2020} FARIAS M. B.\textit{et al.}, \textit{npj Quantum Inf}, \textbf{6} (2020) 25.

 \bibitem{Lombardo2021} LOMBARDO F. C.. \textit{et al.}, \textit{Adv. Quant. Tech}, \textbf{4} (2021) 2000155.
 
\bibitem{Laliotis21} LALIOTIS A. \textit{et al.}, \textit{AVS Quantum Sci.}, {\bf 3} (2021) 043501.

\bibitem{Shresta2003} SHRESTA S., HU B.-L. and PHILLIPS N. G., \textit{Phys. Rev. A}, {\bf 68} (2003) 062101.

\bibitem{Vasile2008} VASILE R. and PASSANTE R.,  \textit{Phys. Rev. A}, {\bf 78} (2008) 032108. 

\bibitem{Messina2010} MESSINA R., VASILE R. and PASSANTE R.,  \textit{Phys. Rev. A}, {\bf 82} (2010) 062501. 

\bibitem{Behunin2011} BEHUNIN R. and HU B.-L.,  \textit{Phys. Rev. A}, {\bf 84} (2011) 012902.

\bibitem{Antezza2014} ANTEZZA M. \textit{et al.}, \textit{Phys. Rev. Lett.} {\bf 113} (2014) 023601.

\bibitem{Perreault2005}  PERREAULT J.D. and CRONIN A. D., \textit{Phys. Rev. Lett.}, \textbf{95} (2005) 133201.

\bibitem{Vigue09} LEPOUTRE S. \textit{et al.}, \textit{Eur. Phys. Lett.}, \textbf{88} (2009) 20002.

\bibitem{Vigue11} LEPOUTRE S. \textit{et al.}, \textit{Eur. Phys. J. D}, \textbf{62} (2011) 309.

\bibitem{Bender14}  BENDER H \textit{et al.},  \textit{Phys. Rev. X}, 4 (2014) 011029.

\bibitem{Pancharatnam56} PANCHARATNAM S., \textit{Proc. Ind. Acad. Science A}, \textbf{44} (1956) 247.

\bibitem{Berry84} BERRY M. V., \textit{Proc. R. Soc. Lond. A}, \textbf{392} (1984) 45.

\bibitem{thiru} CRAIG D. P. and THIRUNAMACHANDRAN T., \textit{Molecular quantum electrodynamics} (Academic Press, London) 1998.

\bibitem{passante21} PASSANTE R. and RIZZUTO L., \textit{Symmetry}, {\bf 13} (2021) 2375.

\bibitem{andrews18} ANDREWS D. L.  \textit{et al.},  \textit{J. Chem. Phys.}, {\bf 148} (2018) 040901. 

\bibitem{Baxter93} BAXTER C.,  BABIKER M. and LOUDON R., \textit{Phys. Rev.  A}, {\bf 47} (1993) 1278.	

\bibitem{Wilkens94} WILKENS M.,  \textit{Phys. Rev. A}, {\bf 49} (1994) 570.

\bibitem{Grimm00} GRIMM R., WEIDEM\"ULLER M., and OVCHINNIKOV Y. B., \textit{Adv. Atom. Mol. Opt. Phys.}, {\bf 42} (2000) 95. 

\bibitem{Amico21} AMICO L.  \textit{et al.}, \textit{AVS Quantum Sci.}, {\bf 3} (2021) 039201.

\bibitem{Passante98} PASSANTE R., POWER E. A. and THIRUNAMACHANDRAN T., \textit{Phys. Lett. A}, {\bf 249}
(1998) 77.

\bibitem{Impens13a}  IMPENS F. \textit{et al.}, \textit{Eur. Phys. Lett.}, \textbf{101} (2013) 60006.

\bibitem{Impens14} IMPENS F. \textit{et al.}, \textit{Phys. Rev. A}, \textbf{89} (2014) 022516.

\bibitem{Impens13b} IMPENS F., TTIRA C. C. and MAIA NETO P. A., \textit{J. Phys. B: At. Mol. Opt. Phys.}, \textbf{46} (2013) 245503.

\bibitem{Antezza2005} ANTEZZA M., PITAEVSKII L. P. and STRINGARI S., \textit{Phys. Rev. Lett.}, {\bf 95} (2005) 113202. 

\bibitem{Obrecht2007} OBRECHT J. M. \textit{et al.}, \textit{Phys. Rev. Lett.}, {\bf 98} (2007) 063201.

\bibitem{Buhmann2008}  BUHMANN S. Y. and SCHEEL S.,  \textit{Phys. Rev. Lett.}, {\bf 100} (2008)  253201.

\bibitem{Behunin2010} BEHUNIN R. O. and HU B.-L.,  \textit{J. Phys. A: Math. Theor.}, {\bf 43} (2010) 012001.

\bibitem{Bartolo2016} BARTOLO N.  \textit{et al.}, \textit{Phys. Rev. A}, {\bf 93} (2016) 042111. 

\bibitem{Matos21} MATOS G. C. \textit{et al.}, \textit{Phys. Rev. Lett.}, \textbf{127} (2021) 270401.

\bibitem{Miniatura92} MINIATURA Ch. \textit{et al.}, \textit{Phys. Rev. Lett.}, \textbf{69} (1992) 261. 

\bibitem{Lepoutre12} LEPOUTRE S. \textit{et al.}, \textit{Phys. Rev. Lett.}, {\bf 109} (2012) 120404.

\bibitem{Gillot13} GILLOT J. \textit{et al.}, \textit{Phys. Rev. Lett.}, {\bf 111} (2013)  030401.

\bibitem{Shapere1989} SHEPERE A. and WILCZEK F., \textit{Geometric Phases in Physics}
(World Scientific, Singapore) 1989.

\bibitem{Borde01} BORD\'E Ch. J., \textit{C. R. Acad. Sci. Paris, Ser. IV}, {\bf 2} (2001) 509.

\bibitem{Impens09} IMPENS F. and BORD\'E Ch. J., \textit{Phys. Rev. A}, {\bf 79} (2009) 043613.

\bibitem{Aharonov13} AHARONOV Y. \textit{et al.}, \textit{New J. Phys.}, \textbf{15} (2013) 113015.

\bibitem{Das20} DAS D. and PATI A. K., \textit{New J. Phys.} \textbf{22}  (2020)  063032.

\bibitem{Liu20} LIU Z.-H. \textit{et al.}, \textit{Nat. Commun.}, \textbf{11} (2020) 3006.

\bibitem{CeshireCat21} KIM Y. \textit{et al.}, \textit{npj quant. inf.}, \textbf{7} (2021) 13.

\bibitem{Souza2016} DE MELO E SOUZA R., IMPENS F. and MAIA NETO P. A., \textit{Phys. Rev. A}, {\bf 94} (2016) 062114.
 
\bibitem{Feynmann63} FEYNMAN R. P. and VERNON F. L., \textit{Ann. Phys. (N.Y.)}, \textbf{24} 118 (1963).

\bibitem{Manjavacas10} MANJAVACAS A. and GARC\'IA DE ABAJO F. J.,  \textit{Phys. Rev. A}, {\bf  82} (2010) 063827. 

\bibitem{Aharonov1959} AHARONOV Y.  and BOHM D.,  
  \textit{Phys. Rev.}
{\bf 115} (1959) 485. 

\bibitem{Dalibard11} DALIBARD J. \textit{et al.}, \textit{Rev. Mod. Phys.}, \textbf{83} (2011) 1523.

\bibitem{Paiva21} PAIVA I. L. \textit{et al.}, \textit{e-print}  arXiv:2110.05824 (2021).

\bibitem{Sciama1953} SCIAMA D., \textit{Mon. Not. R. Astron. Soc.}, {\bf 113} (1953) 34. 

\bibitem{Delic2020} DELI\'C, U. \textit{et al.}, \textit{Science}, {\bf 367} (2020) 892.

\bibitem{Tebbenjohanns2021} TEBBENJOHANNS F.  \textit{et al.}, \textit{Nature}, {\bf 595} (2021) 378. 

\bibitem{Stickler2021} STICKLER B. A., HORNBERGER K. and KIM M. S.,  \textit{Nat. Rev. Phys.}, {\bf 3} (2021)  589.

\end{thebibliography}
\end{document}